\documentclass[twocolumn,showpacs,amsmath,amssymb,showkeys,floatfix,prl]{revtex4}
\usepackage{amsfonts,amsmath}
\usepackage{graphicx}
\usepackage{dcolumn}
\usepackage{bm}

\begin{document}

\title{Bounds on the angles for the parameterization of three neutrino mixing}

\author{D. C. Latimer and D. J. Ernst}

\affiliation{Department of Physics and Astronomy, Vanderbilt University, 
Nashville, Tennessee 37235, USA}
\date{\today}

\begin{abstract}
An algebraic approach is used to derive symmetry relations for the parameterization of the three neutrino mixing matrix.
 The symmetry relations imply bounds on the mixing angles. Including a CP violating phase $\delta$, the mixing angles
$\theta_{jk}$ lie in the range [0,$\pi$/2], and $\delta$ lies within [0,$2\pi$).  If one restricts the CP phase $\delta \in [0,\pi)$,
then one must extend the range of either $\theta_{12}$ {\it or} $\theta_{23}$ to $[0, \pi)$. In particular, for no CP
violation, one can set $\delta=0$ {\it and} $\delta=\pi$ with $\theta_{jk}$ in the range [0,$\pi$/2].
Alternatively, one can set $\delta=0$ only, and allow either $\theta_{12}$ {\it or} $\theta_{23}$ to lie within 
[0,$\pi$).
\end{abstract}



\maketitle

An abundance of data now demonstrates that neutrinos oscillate between flavor states while propagating in vacuum.
Analyses of the data in the context of three neutrino mixing can be found in 
Refs.~\cite{der,fog0,fog1,fog2,ahl1,wei,ahl2,ohl1,gon1,ahl3,gon3,fog4,fog3,gon2,lat1,gon4,mal,ble}.
Conventionally, the standard model of the electroweak 
interaction is extended to include a mass term for the neutrinos and to include a unitary mixing matrix 
which relates the flavor eigenstates of the neutrinos to the mass states of the neutrinos.
The transformation is given by the unitary transformation $U_{\alpha j}$,
\begin{equation}
\nu_\alpha = \sum_j U_{\alpha j} \nu_j\,\,,
\end{equation}
where we use Roman letters, $j$, for mass eigenstates and Greek letters, $\alpha$, for 
flavor eigenstates.

For only two neutrinos, the mixing
matrix reduces to an element of the commutative group $U(1)$. The symmetries 
of this theory are readily apparent as vacuum oscillations 
depend only on one mixing angle and one mass-squared difference, in the limit of $p\gg m$. 
In a three neutrino theory, the mixing matrix is an element of a non-commutative group.
Symmetries in the parameterization are then less obvious. We here derive, for propagation in 
vacuum, symmetries among the mixing angles
and the bounds on the mixing angles which they imply. It is important in analyzing data 
that the full space of allowed mixing angles be explored. The limits on mixing angles for
for the progation through matter have previously been considered in \cite{gluza}.  Our 
results for vaccum propagation are in agreement with theirs;
however, our use of the symmetry relations permits us to arrive at an additional 
alternative conclusion.
A discussion of the four neutrino 
case can be found in Ref.~\cite{pas}.

Casting the theory in algebraic terms allows us
to determine symmetries of the oscillation probability most easily.
We refer to \cite{kchdcl1} for an algebraic approach to vacuum neutrino
oscillations.  The oscillation probability, valid for an arbitrary
fixed number of neutrino mass eigenstates, is
\begin{equation}
\mathcal{P}_{\alpha \to \beta}(t) =\frac{1}{2} \mathrm{tr}[P_+
e^{iHt}  P^\alpha
e^{-iHt} P_+ P^\beta ]\,\,,
\label{mixa}
\end{equation}
where $P^\alpha$ is the flavor projection operator given by $(P^\alpha)_{jk}
= (U^\dagger E^\alpha U)_{jk}= U^*_{\alpha j} U_{\alpha k}$ 
in the mass-eigenstate basis, $P_+$ projects onto
the positive-energy states, and $H$ is an $n$-particle Dirac Hamiltonian for
masses $m_j$. We note that adding multiples of the identity to the
Hamiltonian leaves Eq.~(\ref{mixa}) invariant and recall that
the trace is cyclical.  
For the case of three neutrinos, the relativistic
limit of the above equation is equivalent to the usual oscillation formula
\begin{subequations}
\begin{eqnarray}
\mathcal{P}_{\alpha \to \beta}(L/E)&=& \mathrm{tr}
[U e^{i\mathcal{M}L/2E} U^\dagger E^\alpha U e^{-i\mathcal{M}L/2E} U^\dagger E^\beta] \nonumber \\ 
\label{osctrace} \\ 
&=& \delta_{\alpha \beta}\nonumber\\
&&-4 \sum^3_{\genfrac{}{}{0pt}{}{j <
k}{j,k=1}} \mathrm{Re} (U_{\alpha j} U^*_{\alpha k} U_{\beta k} 
U^*_{\beta
j}) \sin^2 (\varphi_{jk})
\nonumber \\ 
&& +2 \sum^3_{\genfrac{}{}{0pt}{}{j < k}{j,k=1}} \mathrm{Im} (U_{\alpha
j} U^*_{\alpha k} U_{\beta k}
U^*_{\beta j}) \sin(2 \varphi_{jk}) \nonumber\\
\label{oscform}
\end{eqnarray}
\end{subequations}
where $\mathcal{M}$ is the diagonal matrix with entries $(m_1^2, m_2^2, m_3^2)$ and
$\varphi_{jk} := \Delta_{jk} L/4E$ with $\Delta_{jk} := m_j^2 -
m_k^2$. 

The Maiani parameterization \cite{chau} of the three-neutrino
mixing matrix is
\begin{widetext}
\begin{equation}
U(\theta_{23},\theta_{13},\theta_{12},\delta) = \left(  
\begin{array}{ccc}
c_{12} c_{13} & s_{12} c_{13} & s_{13} e^{-i \delta} \\
-s_{12}c_{23} - c_{12} s_{23} s_{13} e^{i \delta}  & 
c_{12} c_{23} - s_{12} s_{23} s_{13}  e^{i \delta}& s_{23}c_{13}\\
 s_{12}s_{23} - c_{12} c_{23} s_{13} e^{i \delta} & 
-c_{12} s_{23} - s_{12} c_{23} s_{13}  e^{i \delta}& c_{23}c_{13}
\end{array}
\right)\,\,, \label{mixer}
\end{equation}
\end{widetext}
where $c_{jk} = \cos{\theta_{jk}}$,
$s_{jk}=\sin{\theta_{jk}}$, and $\theta_{jk}$, $\delta$ are real.
Noting that the mass-squared differences must satisfy
\begin{equation}
\Delta_{12} + \Delta_{23} +\Delta_{31} = 0\,\,,
\end{equation}
the oscillation probability in Eqs.~(3) contains six
independent parameters.  We assemble the mixing angles and CP phase 
of Eq.~(\ref{mixer}) into the quadruple 
$(\theta_{23}, \theta_{13}, \theta_{12},\delta)$ with the understanding 
that the mass-squared differences remain unchanged.  

Invariance of the oscillation probability in (\ref{oscform}) is 
equivalent to the invariance of the trace in (\ref{osctrace}). 
We employ the fact that the trace
of an operator $A$ remains unchanged when conjugated by any unitary 
operator $U$
\begin{equation}
\mathrm{tr}[A] = \mathrm{tr}[UAU^\dagger]\,\,. \label{trprop}
\end{equation} 
We consider only parameter changes that leave the mass-squared differences
fixed. We use Eqs.~ (3)
to define an equivalence class among parameter sets;
explicitly, one set of parameters is deemed equivalent to another set of parameters 
if the oscillation probabilities
$P_{\alpha \to \beta}(L/E)$ are identical for all values of $\alpha$ and $\beta$.  
This equivalence will be expressed via the relation $\equiv$.  

The group $SO(3)$ has three generators. We shall 
represent the group as real matrices acting on $\mathbb{R}^3$
and express the exponentiated generators as $U_j(\theta)$, 
a rotation by angle $\theta$ about the $j$th
axis.  We need make the additional definition for $S_\delta \in SU(3)$ by setting 
$S_\delta = \mathrm{diag} (e^{-i\delta/2},1,e^{i\delta/2})$ and letting
\begin{equation}
U_2^\delta(\theta)= S_\delta U_2(\theta) S_\delta^\dagger\,\,. \label{u2delta}
\end{equation}
The neutrino mixing matrix (\ref{mixer}) may be expressed as
\begin{equation}
U(\theta_{23},\theta_{13},\theta_{12},\delta)= 
U_1(\theta_{23}) U_2^\delta(\theta_{13}) U_3(\theta_{12})\,\,.
\end{equation}
The generalization to $N$ neutrinos may be found in Ref.~\cite{gro}.

From Eq.~(\ref{u2delta}), we may extract our first equivalence relation.  
For the CP phase $\delta=\pi$, one finds
\begin{equation}
S_\pi U_2(\theta) S_\pi^\dagger = U_2(-\theta)\,\,;
\end{equation}
hence, one has
\begin{equation}
U_2^{\delta+\pi}(\theta)=U_2^\delta(-\theta)\,\,.
\end{equation}
The addition of $\pi$ to the CP phase can accommodate a change in sign of $\theta_{13}$
\begin{equation}
(\theta_{23}, \theta_{13}, \theta_{12}, \delta) \equiv  (\theta_{23}, -\theta_{13}, 
\theta_{12}, \delta +\pi)\,\,. \label{deltapi}
\end{equation}

Proceeding to further symmetries, a simple calculation confirms 
\begin{equation}
U_j(\pi)U_k(\theta)U_j(\pi)^\dagger = U_k(-\theta) \label{picommute}
\end{equation}
for $j \ne k$.  Additionally, as $U_j(\pi)$ is diagonal, it commutes
with $S_\delta$ so that
\begin{equation}
U_j(\pi)U_2^\delta(\theta)U_j(\pi)^\dagger = U_2^\delta(-\theta)
\end{equation}
for $j=1,3$.  For the same reason, $U_j(\pi)$ commutes with
the mass-squared matrix $\mathcal{M}$ and the projection $E^\alpha$.  
Combining these results with Eq.~(\ref{trprop}), we note that the following
parameters yield equivalent oscillation probabilities for all values of $\delta$
\begin{subequations} 
\begin{eqnarray}
(\theta_{23}, \theta_{13}, \theta_{12}, \delta) &\equiv&
(-\theta_{23}, -\theta_{13}, \theta_{12}, \delta) \label{minus1}\\
&\equiv& (-\theta_{23}, \theta_{13}, -\theta_{12}, \delta) \label{minus2} \\ 
&\equiv& (\theta_{23}, -\theta_{13}, -\theta_{12}, \delta)\,\,. \label{minus3} 
\end{eqnarray}
\end{subequations}
Additionally, one can determine the effect
of adding $\pi$ to a mixing angle; one finds
\begin{subequations}
\begin{eqnarray}
(\theta_{23}, \theta_{13}, \theta_{12}, \delta) &\equiv&
(\theta_{23} + \pi, \theta_{13}, \theta_{12},\delta) \label{pi1}\\ 
&\equiv& (-\theta_{23}, \theta_{13} + \pi, \theta_{12},\delta) \label{pi2}\\
&\equiv& (\theta_{23}, \theta_{13},\theta_{12} + \pi,\delta)\,\,. \label{pi3}
\end{eqnarray}
\end{subequations}

Trivially, the mixing angles satisfy a $2\pi$ periodicity; however, from
the relations in (\ref{pi1}--\ref{pi3}), it is clear that, without loss of generality, 
one may further restrict all $\theta_{jk}$ 
to lie \cite{ftn} in the interval $[0,\pi)$.  In the case of $\theta_{13}$, we are able to further
narrow these bounds to $[0,\pi/2]$.  An application of relation
(\ref{minus1}) followed by (\ref{pi2}) yields 
\begin{equation}
(\theta_{23}, \theta_{13}, \theta_{12}, \delta) \equiv
(\theta_{23} , \pi - \theta_{13}, \theta_{12},\delta)\,\,. \label{pihalf13}
\end{equation}
In general, given a real number $x$, the map \mbox{$x \mapsto a - x$} is a reflection 
about the point $a/2$ on the real line.  Hence, we see that if 
$\theta_{13}$ should lie between $\pi/2$ and $\pi$, then the above relation shows that
we have an equivalent oscillation probability for an angle reflected about $\pi/2$.  
In short, we may choose $\theta_{13}$ to lie in the first quadrant.  

For the remaining mixing angles, we may apply relations (\ref{minus2}), (\ref{pi1}), 
and (\ref{pi3}) to achieve
\begin{equation}
(\theta_{23}, \theta_{13}, \theta_{12}, \delta) \equiv
(\pi- \theta_{23} , \theta_{13}, \pi-\theta_{12},\delta)\,\,. \label{pihalf23}
\end{equation} 
From this, one notes that a reflection about $\pi/2$ in both $\theta_{23}$ 
and $\theta_{12}$ results in an equivalent oscillation probability.  These reflections
cannot be performed independently and still result in a equivalent theory.  Hence, we only 
have the freedom to restrict either $\theta_{23}$ or $\theta_{12}$ to the interval $[0,\pi/2]$, 
but not both.

Relation (\ref{pihalf23}) is valid for fixed $\delta$.  Should we relax this condition on $\delta$,
then we have the liberty to restrict both mixing angles to lie between $0$ and $\pi/2$.  
Using (\ref{deltapi}), (\ref{minus1}), and (\ref{pi1}), we have
\begin{equation}
(\theta_{23}, \theta_{13}, \theta_{12}, \delta) \equiv
(\pi- \theta_{23} , \theta_{13}, \theta_{12},\delta + \pi)\,\,.\label{lastm1}
\end{equation}
From similar logic, one can also deduce
\begin{equation}
(\theta_{23}, \theta_{13}, \theta_{12}, \delta) \equiv
(\theta_{23} , \theta_{13}, \pi - \theta_{12},\delta + \pi)\,\,.\label{last}
\end{equation}
Permitting a change in the CP phase results in the ability to independently reflect these two mixing
angles about $\pi/2$; hence, we are guaranteed, as noted in \cite{gluza, har}, that all $\theta_{jk}$ can be restricted to the
interval $[0,\pi/2]$ if one allows the full range on the CP phase $\delta \in [0, 2\pi)$.  This is the common
understanding for the CKM mixing matrix for quarks.  

As a result of relation (\ref{deltapi}), the full physics of CP violation in neutrino oscillations can be realized by
restricting the CP phase to lie within the smaller interval $\delta \in [0,\pi)$.  This is a tacit assumption of many
neutrino phenomenologists as they often work in the limit of no CP violation by setting $\delta=0$.  As indicated in
\cite{gluza}, this restriction does not account for all the necessary parameter space if all mixing angles are in the
first quadrant.  We show that if one does indeed choose to restrict
the phase to this interval, then one must compensate for this by extending the range of either mixing angle 
$\theta_{12}$ {\it or} $\theta_{23}$ to $[0, \pi)$.  Hence, for the often studied case of no CP
violation in which we choose a unique value for the CP phase, say $\delta=0$, then we cannot utilize Eqs.~(\ref{lastm1},
\ref{last}), and we must 
increase the range on $\theta_{12}$ {\it or} $\theta_{23}$ to $[0,\pi)$ in order to consider all possible oscillation
scenarios. Alternatively, one could use these relations by including both $\delta=0$ and $\delta = \pi$
and restrict all $\theta_{ij}$ to $[0,\pi/2]$; this is the solution proposed in \cite{gluza}. 
To ensure that we have not overlooked a symmetry which could further reduce the bounds
on the mixing angles, we have checked numerically that the mixing probabilities are indeed 
unique over the regions given.

\bibliography{angles}

\begin{thebibliography}{1}

\bibitem{der}
A. de R\'{u}jula, M. Lusignoli, L. Maiani, S. T. Petcov, and 
R. Petronzio,
Nucl. Phys. {\bf B168}, 54 (1980).

\bibitem{fog0}
G. L. Fogli, E. Lisi, and D. Montanino, Phys. Rev. D {\bf 49},
3626 (1994); {\bf 54}, 2048 (1996); {\bf 64}, 093005 (2001)

\bibitem{fog1}
G. L. Fogli, E. Lisi, and G. Scioscia, Phys. Rev. D {\bf 52},
5334 (1995).

\bibitem{fog2}
G. L. Fogli, E. Lisi, D. Montanino, 
and G. Scioscia, Phys. Rev. D {\bf 55}, 4385 (1997).

\bibitem{ahl1}
D. V. Ahluwalia, Mod. Phys. Lett. A {\bf 13}, 2249 (1998).

\bibitem{wei}
V. Barger, S. Pakvasa, T. J. Weiler, and K. Whisnant, Phys. Lett. 
{\bf B437}, 107 (1998).

\bibitem{ahl2}
I. Stancu and D. V. Ahluwalia, Phys. Lett. {\bf B460}, 431 (1999).

\bibitem{ohl1}
T. Ohlsson and H. Snellman, Phys. Rev. D {\bf 60}, 093007 (1999).

\bibitem{gon1}
M. C. Gonzalez-Garcia, M. Maltoni, C. Pe\~{n}a-Garay, and J. W. F.
Valle, Phys. Rev. D {\bf 63}, 033005 (2001).

\bibitem{ahl3}
D. V. Ahluwalia, Y. Liu, and I. Stancu, Mod. Phys. Lett. A {\bf 17},
13 (2002).

\bibitem{gon3}
M.C. Gonzalez-Garcia and M. Maltoni, Eur. Phys. J. C {\bf 26}, 417 (2003).

\bibitem{fog4}
G. L. Fogli, E. Lisi, D. Montanino, and A. Palazzo, Phys. Rev. D
{\bf 65}, 073008 (2002).

\bibitem{fog3}
G. L. Fogli, G. Lettera, E. Lisi, A. Marrone, A. Palazzo, and A. Rotunno,
Phys. Rev. D {\bf 66}, 093008 (2002).

\bibitem{gon2}
M. C. Gonzalez-Garcia and Y. Nir, Rev. Mod. Phys. {\bf 75}, 345 (2003).

\bibitem{lat1}
D. C. Latimer and D. J. Ernst, nucl-th/03100830.

\bibitem{gon4}
M. C. Gonzalez-Garcia and C. Pe\~na-Garay, Phys. Rev. D {\bf 68}, 09003 (2003).

\bibitem{mal}
M. Maltoni, T. Schwartz, M. A. T\'{o}rtala, and J. W. F. Valle, Phys.
Rev. D {\bf 68}, 113010 (2003). 

\bibitem{ble}
M. Blennow, T. Ohlsson, and H. Snellman, hep-th/0311098.

\bibitem{gluza}
J. Gluza and M. Zralek, Phys. Lett. {\bf B517}, 158 (2001).

\bibitem{pas}
H. P\"{a}s, L. Song, and T. J. Weiler, Phys. Rev. D {\bf 67}, 073019 (2003).

\bibitem{kchdcl1}
K. C. Hannabuss and D. C. Latimer, J. Phys. A: Math. Gen. {\bf 33},
1369 (2000).

\bibitem{chau}
M. Maiani, Phys. Lett. {\bf B62}, 183 (1976).

\bibitem{gro}
M. Gronau, R. Johnson, and J. Schecter, Phys. Rev. D {\bf 32}, 3062 
(1985).

\bibitem{ftn} We utilize the curved bracket $)$ to indicate that the upper limit
is not included within the range and utilize the square bracket ] when the limit 
is a part of the range.
 

\bibitem{har}
H. Harari and M. Leurer, Phys. Lett. {\bf B181}, 123 (1986).

\end{thebibliography}

\end{document}